\documentclass[a4paper]{jpconf}

\usepackage{array}
\usepackage{braket}
\usepackage{slashed}
\usepackage{graphicx}
\usepackage{subfig}
\usepackage{amsmath, amssymb, amsfonts}
\usepackage{mathtools}
\usepackage{mathrsfs}
  \DeclareMathAlphabet{\mathpzc}{OT1}{pzc}{m}{it}
\usepackage{bm}

\usepackage{textpos}
\usepackage{etoolbox}
\patchcmd{\thebibliography}{\advance\leftmargin\labelsep}
  {\labelsep=0.5cm \advance\leftmargin\labelsep}{}{}
  
\begin{document}
\title{Self-consistent radiative corrections to\\ false vacuum decay}
\author{
B Garbrecht\textsuperscript{1} and \underline{P Millington}\textsuperscript{2}}
\address{\textsuperscript{1} Physik Department T70, Technische Universit\"{a}t M\"{u}nchen\\ James-Franck-Stra\ss e, 85748 Garching, Germany\\
  \textsuperscript{2}School of Physics and Astronomy, University of Nottingham,\\ Nottingham NG7 2RD, United Kingdom}
\ead{garbrecht@tum.de, p.millington@nottingham.ac.uk}

\begin{textblock}{4}(12,-9.2)
\begin{flushright}
\begin{footnotesize}
TUM/HEP/1076/17 \\
15 March 2017
\end{footnotesize}
\end{flushright}
\end{textblock}

\begin{abstract}
With the Higgs mass now measured at the sub-percent level, the potential metastability of the electroweak vacuum of the Standard Model (SM) motivates renewed study of false vacuum decay in quantum field theory. In this note, we describe an approach to calculating quantum corrections to the decay rate of false vacua that is able to account fully and self-consistently for the underlying inhomogeneity of the solitonic tunneling configuration. We show that this method can be applied both to theories in which the instability arises already at the level of the classical potential and those in which the instability arises entirely through radiative effects, as is the case for the SM Higgs vacuum. We analyse two simple models in the thin-wall regime, and we show that the modifications of the one-loop corrections from accounting fully for the inhomogeneity can compete at the same level as the two-loop homogeneous corrections.
\end{abstract}

\section{Introduction}

The dynamics of first-order phase transitions have relevance across high-energy physics, astro-particle physics and cosmology. Given a $\sim 125\ {\rm GeV}$ Higgs boson and a $\sim 174\ {\rm GeV}$ top quark, state-of-the-art perturbative calculations of the Standard Model (SM) effective potential indicate that the electroweak vacuum may suffer an instability~\cite{EliasMiro:2011aa,Degrassi:2012ry,Alekhin:2012py,Buttazzo:2013uya,Bednyakov:2015sca,DiLuzio:2015iua} at a scale of around $10^{11}\ {\rm GeV}$, leading to a catastrophic transition towards a high-scale global minimum that lies far above the Planck scale. In the cosmology of the early Universe, first-order thermal phase transitions may play a role in the generation of the Baryon Asymmetry of the Universe~\cite{Morrissey:2012db,Chung:2012vg} and have the potential to produce stochastic gravitational-wave backgrounds~\cite{Witten:1984rs,Kosowsky:1991ua,Kosowsky:1992rz,Caprini:2009fx}. Further study of the latter is certainly warranted given the recent discovery of gravitational waves by LIGO~\cite{LIGO} and the potential for such a stochastic background to be observed by the forthcoming LISA satellite array or pulsar-timing arrays. Moreover, in analysing vacuum decay, we must study non-perturbative solutions of non-linear field equations, and the necessary techniques are also of relevance to the formation and evolution of (non-)topological defects.

In the case of the SM electroweak vacuum, the predicted lifetime is subject to a number of uncertainties. The main experimental uncertainty is due to the determination of the top-quark mass~\cite{Bezrukov:2012sa,Masina:2012tz}. More significantly, the exit point for the tunneling can be close to the Planck scale, and one therefore cannot neglect the impact of new physics and higher-dimension (non-renormalisable) operators~\cite{Branchina:2013jra,Branchina:2014usa,Lalak:2014qua,Branchina:2014rva,Eichhorn:2015kea}, which can significantly impact the tunneling rate. Moreover, one should consider the effect of gravitational corrections~\cite{Branchina:2016bws,Czerwinska:2016fky,Burda:2015isa,Rajantie:2016hkj}, as well as the role played by ``seeded'' decay, which can result from impurities~\cite{Grinstein:2015jda} due, for instance, to black holes~\cite{Gregory:2013hja,Burda:2015yfa,Burda:2016mou}. On the theoretical side, we must be careful about the interpretation of non-convex regions in the perturbatively-calculated effective potential~\cite{Weinberg:1987vp,Alexandre:2007ci,Branchina:2008pc}, as well as the implementation of the renormalisation-group (RG) improvement of the effective potential~\cite{Einhorn:2007rv,Gies:2014xha}. In this note, we will concentrate on the additional uncertainty introduced by neglecting the spatial inhomogeneity of the underlying solitonic field configuration. This inhomogeneity has not been captured fully in existing analyses of the SM vacuum stability, where the tunneling rate is calculated using the effective potential of a homogeneous field configuration. To this end, we will outline a Green's function method, based on Refs.~\cite{Garbrecht:2015oea,Garbrecht:2015cla,Garbrecht:2015yza}, that allows quantum-corrected tunneling rates to be calculated self-consistently within the space-time dependent background of the tunneling configuration.

We begin by reviewing the semi-classical calculation of the tunneling rate, before discussing the calculation of the first quantum corrections. We then apply the Green's function method to two toy models. In the first model (reported from Ref.~\cite{Garbrecht:2015oea}), the vacuum instability is present at tree level, and we will calculate the quantum corrections analytically in the inhomogeneous background field configuration. In the second model (reported from Ref.~\cite{Garbrecht:2015yza}), the vacuum instability emerges only at the one-loop level, and, therein, we will apply a self-consistent numerical procedure to calculate the quantum-corrected tunneling rate. We will find that the modifications to the one-loop results from fully accounting for the background inhomogeneity can compete at the same level as the two-loop (homogeneous) corrections.

\section{Semi-classical tunneling rate}

We begin by reviewing the calculation of the semi-classical tunneling rate for the archetypal quartic theory with Euclidean Lagrangian
\begin{subequations}
\label{eq:Lagrangian}
\begin{gather}
\mathcal{L} \ = \ \frac{1}{2!}\,\big(\partial_{\alpha}\Phi\big)^2\:+\:U(\Phi)\;,\\
U(\Phi)\ =\ -\:\frac{1}{2!}\,\mu^2\,\Phi^2\:+\:\frac{1}{3!}\,g\,\Phi^3\:+\:\frac{1}{4!}\,\lambda\,\Phi^4\:+\:U_0\;.
\end{gather}
\end{subequations}
Herein, $\Phi\equiv\Phi(x)$ is a real scalar field, $\partial_{\alpha}$ indicates the partial derivative with respect to the Euclidean four-coordinate $x_{\alpha}\equiv (\mathbf{x},x_4)$, $g,\lambda>0$ are dimensionful and dimensionless couplings, respectively, and $U_0$ is a constant. For $\mu^2>0$, the system undergoes spontaneous symmetry breaking, and the field obtains a vacuum expectation value $\varphi\equiv\braket{\Phi}=v_{\pm}\simeq \pm\,v-3g/(2\lambda)$, where $v=\sqrt{6}\mu/\sqrt{\lambda}$. The cubic coupling $g$, which we take to be small ($3g/(2\lambda)\ll v$), breaks the $\mathbb{Z}_2$ symmetry, such that the minima at $v_{\pm}$ are not fully degenerate. The calculation that follows was first presented by Coleman in Ref.~\cite{Coleman:1977py}, based on the work of Langer~\cite{Langer:1969bc} on the statistical description of the decay of metastable states.

\begin{figure}
\centering
\includegraphics[scale=1]{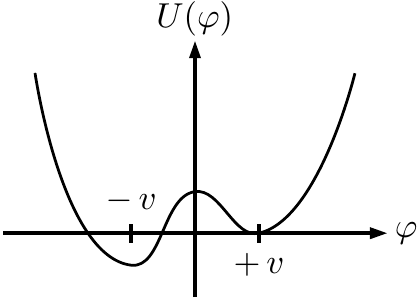}\hspace{2cm}\includegraphics[scale=1]{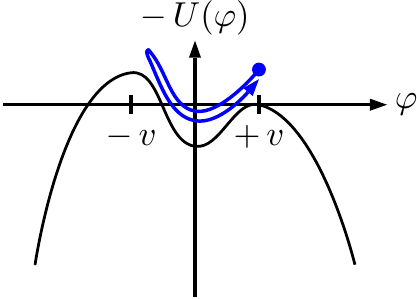}
\caption{\label{fig:treepotential}Sketches of the tree-level potential (left) and the upturned potential (right) for the model in Eq.~\eqref{eq:Lagrangian}. The blue line (right) indicates the trajectory of the bounce.}
\end{figure}

An initially homogeneous state of the false vacuum ($\varphi=+\,v$) will spontaneously fragment via the nucleation of bubbles, which then expand and coalesce to fill the space with the true vacuum ($\varphi=-\,v$). The rate of nucleation of such bubbles depends upon the difference in the free energies of the two vacuum states, and the tunneling field configuration corresponds to the lowest intervening saddle point of the free energy. Since the free energy is related to the Euclidean action, we are interested in the classical Euclidean equation of motion
\begin{equation}
\label{eq:eom}
-\:\Delta^{(4)}\,\varphi\:+\:U'(\varphi)\ =\ 0\;,
\end{equation}
where $\Delta^{(4)}$ is the four-dimensional Laplace-Beltrami operator and $\prime$ denotes differentiation with respect to the field $\varphi$. The Euclidean equation of motion is equivalent to that of a classical particle moving in an upturned potential $-\,U(\varphi)$ (see Fig.~\ref{fig:treepotential}), and the tunneling corresponds to the trajectory that starts at the false vacuum in the distant Euclidean `past,' reaches close to the true vacuum at the origin of Euclidean `time,' and rolls back to the false vacuum in the distant `future.' Coleman called this solution \emph{the bounce}, and its boundary conditions are $\varphi(x)|_{x_4\to\pm\,\infty}=+\,v$ and $\partial\varphi(x)/\partial x_4|_{x_4=0}=0$. In order to obtain a finite action, we set $U_0=(\mu v/2)^2-gv^3/6$, so that the false vacuum has zero energy density, and require $\varphi(x)|_{|\mathbf{x}|\to\infty}=+\,v$, i.e.~that the system far from the bubble is unaffected by its nucleation. After transforming to four-dimensional hyperspherical coordinates, the boundary conditions become $\varphi(r)|_{r\to\infty}=+\,v$ and $\partial\varphi(r)/\partial r|_{r=0}=0$, where $r=\sqrt{x\cdot x}$. Since the turning point at $r=0$ occurs close to $\varphi= -\,v$, we see that bounce corresponds to a four-dimensional spherical bubble, separating false vacuum on the outside from true vacuum on the inside. 

\begin{figure}
\centering
\includegraphics[scale=1]{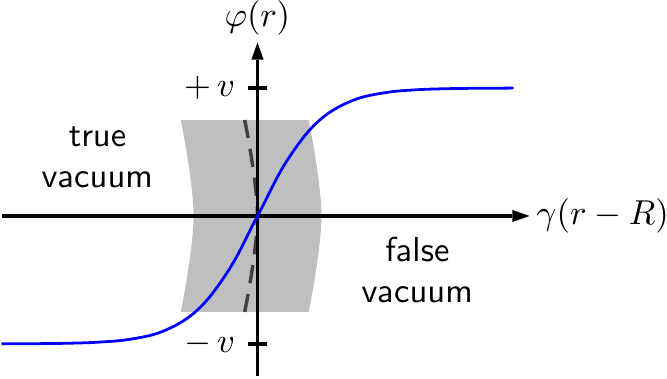}
\caption{\label{fig:kink}Sketch of the classical kink solution [Eq.~\eqref{eq:kink}]. The grey band indicates the position of the bubble wall.}
\end{figure}

In the thin-wall regime, where the diameter of the bubble is much larger than the width of the bubble wall, the gradients of the field are small everywhere except in a narrow region centred on the bubble wall. Since this occurs at some large radius $R\propto 1/g$, we can therefore neglect the damping term $3/r\; {\rm d}\varphi/{\rm d} r$ in the hyperspherical equation of motion
\begin{equation}
\label{eq:hyperradeom}
-\:\frac{{\rm d}^2\varphi}{{\rm d}r^2}\:-\:\frac{3}{r}\,\frac{{\rm d}\varphi}{{\rm d}r}\:+\:U'(\varphi)\ = \ 0\;.
\end{equation}
The relevant solution is the well-known kink (see Fig.~\ref{fig:kink})
\begin{equation}
\label{eq:kink}
\varphi(r)\ =\ v\,\mathrm{tanh}\big[\gamma(r-R)\big]\;,
\end{equation}
where $\gamma\equiv\mu/\sqrt{2}$. To fix the radius of the critical bubble $R$, we extremise the bounce action
\begin{equation}
B\ =\ \int\!{\rm d}^4 x\;\Bigg[\frac{1}{2}\bigg(\frac{{\partial }\varphi}{{\rm \partial}x_{\mu}}\bigg)^{\!2}\:+\:U(\varphi)\Bigg]
\end{equation}
with respect to $R$, giving $R=12\gamma/(gv)$. Wick rotating back to Minkowski spacetime, $x_4=ix_0$, we find that the bubble expands along a hyperbolic trajectory $R^2=|\mathbf{x}|^2\:-\:(ct)^2$, nucleating at $t=0$ with three-dimensional radius $R$ and quickly accelerating to the speed of light ($c$).

The critical bubble actually corresponds to a maximum of the free energy, and the spectrum of the Klein-Gordon operator therefore has a negative eigenvalue
\begin{equation}
\lambda_0\ =\ \frac{1}{B}\,\frac{\delta^2B}{\delta R^2}\ =\ -\,\frac{3}{R^2}\;,
\end{equation}
associated with dilatations of the bounce. As a result, the Euclidean path integral
\begin{equation}
Z[0]\ =\ \int\!\mathcal{D}\Phi\;e^{-\,S[\Phi]/\hbar}
\end{equation}
is ill-defined. Moreover, by virtue of the four-dimensional translational invariance of the action, the spectrum also contains four zero eigenvalues. The integral over these zero modes can be traded for an integral over the collective coordinates of the bounce, and this accounts for the fact that the nucleation of the critical bubble can occur anywhere in the four-volume $VT$. The integral over the negative mode can be performed by the method of steepest decent by deforming the contour of functional integration into the complex plane. As a result, the path integral obtains an imaginary part, which is related to the tunneling rate per unit volume as
\begin{equation}
\label{eq:tunnel1}
\varGamma/V\ =\ 2|\mathrm{Im}\,Z[0]|/(VT)\ \sim\ e^{-B/\hbar}\;.
\end{equation}
Hereafter, we will denote the solution to the classical equation of motion in Eq.~\eqref{eq:kink} by $\varphi^{(0)}$ and the classical bounce action by $B^{(0)}\equiv S[\varphi^{(0)}]$.

\section{Quantum corrections}

In order to find the first quantum corrections (see Refs.~\cite{Callan:1977pt} and \cite{Konoplich:1987yd}) and to fix the proportionality in Eq.~\eqref{eq:tunnel1}, we can perform a saddle-point evaluation of the path integral by expanding around the classical solution, writing \smash{$\Phi=\varphi^{(0)}+\hbar^{1/2}\phi$} and truncating at first order in \smash{$\hbar$}. In this way, we find
\begin{equation}
Z[0]\ =\ -\:\frac{i}{2}\,e^{-B^{(0)}/\hbar}\Bigg|\frac{\lambda_0\,\mathrm{det}^{(5)}G^{-1}(\varphi^{(0)})}{(VT)^2\big(\frac{B^{(0)}}{2\pi\hbar}\big)^{\!4}(4\gamma^2)^5\,\mathrm{det}^{(5)}G^{-1}(v)}\Bigg|^{-1/2}\;,
\end{equation}
where the factors of $(VT)^2\big(\frac{B^{(0)}}{2\pi\hbar}\big)^{\!4}$ have arisen from the integral over the translational zero modes, the superscript ``$(5)$'' indicates that the determinant is over only the positive-definite eigenvalues and $G^{-1}(\varphi^{(0)})$ is the Klein-Gordon or \emph{fluctuation operator}
\begin{equation}
G^{-1}(\varphi^{(0)};x,y)\ \equiv\ \frac{\delta^2 S[\Phi]}{\delta \Phi(x)\delta \Phi(y)}\bigg|_{\Phi\,=\,\varphi^{(0)}}\ =\ \delta^{4}(x-y)\Big(-\:\Delta^{(4)}\:+\:U''(\varphi^{(0)})\Big)\;.
\end{equation}
The factor $(4\gamma^2)^5\,\mathrm{det}^{(5)}G^{-1}(v)$ is introduced for normalisation. Substituting the above result into Eq.~\eqref{eq:tunnel1}, the tunneling rate per unit volume is found to be
\begin{equation}
\varGamma/V\ =\ \bigg(\frac{B^{(0)}}{2\pi\hbar}\bigg)^{\!2}(2\gamma)^5|\lambda_0|^{-1/2}\exp\bigg[-\,\frac{1}{\hbar}\Big(B^{(0)}\:+\:\hbar\,B^{(1)}\Big)\bigg]\;,
\end{equation}
where the one-loop (order $\hbar$) corrections are captured by the \emph{fluctuation determinant}
\begin{equation}
B^{(1)}\ =\ \frac{1}{2}\,\Big(\ln\,\mathrm{det}^{(5)}\,G^{-1}(\varphi^{(0)})\:-\:\ln\,\mathrm{det}^{(5)}\,G^{-1}(v)\Big)\ =\ \frac{1}{2}\,\mathrm{tr}^{(5)}\Big(\ln\,G^{-1}(\varphi^{(0)})\:-\:\ln\,G^{-1}(v)\Big)\;.
\end{equation}
This can be evaluated by a number of approaches, including, for instance, the heat kernel method (see, e.g., Ref.~\cite{Vassilevich:2003xt}), the Gel'fand-Yaglom theorem~\cite{Gelfand:1959nq} and the direct integration method due to Baacke and Junker~\cite{Baacke:1993jr,Baacke:1993aj,Baacke:1994ix}. The latter is particularly well suited to numerical applications.

In the model considered so far [Eq.~\eqref{eq:Lagrangian}], the instability is already present at the level of the classical potential. However, in the case of the SM, the instability emerges only at the quantum level, and the relevant quantum path, viz.~the tunneling trajectory, is therefore non-perturbatively far away from the classical path. To see the significance of this observation, we should consider the spectrum of the fluctuation operator. In the tree-level case, an expansion around the solution to the classical equations of motion leads to a fluctuation operator with a non-positive-definite spectrum. On the other hand, when the instability results from radiative effects, the tree-level fluctuation operator will have a positive-definite spectrum, whilst the negative and zero eigenvalues will appear only at the level of the quantum fluctuation operator. In order to capture the instability consistently, we must therefore perform the saddle-point evaluation of the path integral by expanding around the solutions to the quantum equations of motion, both for the one- and two-point functions. Such a procedure, which we refer to as the \emph{method of external sources}~\cite{Garbrecht:2015cla}, can be formulated rigorously within the nPI effective action formalism. Consider, for illustration, the 2PI effective action~\cite{Cornwall:1974vz}
\begin{equation}
\Gamma[\phi,\Delta]\ =\ -\,\hbar\,\ln\,Z[\mathcal{J},\mathcal{K}]\:+\:\mathcal{J}_x[\phi,\Delta]\phi_x\:+\:\frac{1}{2}\,\mathcal{K}_{xy}[\phi,\Delta]\big(\phi_x\phi_y\:+\:\hbar\,\Delta_{xy}\big)\;.
\end{equation}
By choosing the extremal sources $\mathcal{J}$ and $\mathcal{K}$ such that
\begin{equation}
\frac{\delta S[\Phi]}{\delta \Phi}\bigg|_{\Phi\,=\,\varphi}\:-\:\mathcal{J}_x[\phi,\Delta]\:-\:\mathcal{K}_{xy}[\phi,\Delta]\varphi_y\ =\ \frac{\delta \Gamma[\phi,\Delta]}{\delta \phi_x}\bigg|_{\phi\,=\,\varphi}\ =\ 0
\end{equation}
and $\mathcal{G}^{-1}=G^{-1}-\mathcal{K}$ is the dressed inverse two-point function, we can drive the system along the quantum saddle point, as desired.

\subsection{Tree-level instability}

The tunneling rate per unit volume can be written in terms of the effective action as
\begin{equation}
\varGamma/V\ =\ 2|\mathrm{Im}\,e^{-\,\Gamma/\hbar}|/(VT)\;,
\end{equation}
and, for the model in Eq.~\eqref{eq:Lagrangian}, it is convenient to work at the level of the 1PI effective action:
\begin{equation}
\Gamma[\varphi^{(1)}]\ =\ S[\varphi^{(1)}]\:+\:\frac{i\pi \hbar}{2}\:+\:\frac{\hbar}{2}\,\ln\bigg|\frac{\lambda_0\,\mathrm{det}^{(5)}G^{-1}(\varphi^{(1)})}{\tfrac{1}{4}(VT)^2\big(\tfrac{\mathcal{N}^{-2}}{2\pi\hbar}\big)(4\gamma^2)^5\mathrm{det}^{(5)}G^{-1}(v)}\bigg|\:+\:\mathcal{O}(\hbar^2)\;.
\end{equation}
For further discussion of the normalisation of the zero modes $\mathcal{N}$, see Ref.~\cite{Garbrecht:2015yza}. Writing the quantum-corrected bounce as $\varphi^{(1)}=\varphi^{(0)}+\hbar\,\delta\varphi$, we can expand the 1PI effective action around the classical kink solution $\varphi^{(0)}$ to obtain
\begin{equation}
\varGamma/V\ =\ \bigg(\frac{\mathcal{N}^{-2}}{2\pi\hbar}\bigg)^{\!2}(2\gamma)^5|\lambda_0|^{-1/2}\exp\bigg[-\:\frac{1}{\hbar}\Big(B^{(0)}+\hbar\,B^{(1)}\:+\:\hbar^2\,B^{(2)}\:+\:\hbar^2\,B^{(2)\prime}\Big)\bigg]\;.
\end{equation}
The two-loop 1PR corrections (appearing at order $\hbar^2$) are due to expanding the action of the corrected bounce $S[\varphi^{(1)}]$ and the fluctuation determinant $B^{(1)}[\varphi^{(1)}]$ around $\varphi^{(0)}$:
\begin{equation}
B^{(2)}\ = \ -\:\frac{1}{2}\int{\rm d}^4x\;\varphi^{(0)}(x)\Pi(\varphi^{(0)};x)\delta\varphi(x)\ =\ -\:\frac{1}{2}\,B^{(2)\prime}\;.
\end{equation}

\begin{figure}
\centering
\includegraphics[scale=1,trim={0 1.6cm 0 0},clip]{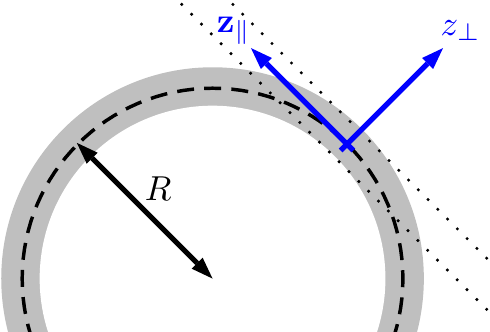}
\caption{\label{fig:planarwall}Sketch of the coordinate system $(z_{\perp},\mathbf{z}_{\parallel})$ applied in the planar-wall limit.}
\end{figure}

Returning to the model in Eq.~\eqref{eq:Lagrangian}, we now wish to calculate the quantum corrections to the tunneling rate, and the starting point will be the Green's function determined within the background of the kink.\footnote{Green's function methods have been applied previously to self-consistent calculations in the Hartree-Fock approximation of the pure $\lambda\Phi^4$ theory in Refs.~\cite{Bergner:2003au,Bergner:2003id, Baacke:2004xk,Baacke:2006kv} and the $\lambda\Phi^4+g\Phi^6$ theory in Ref.~\cite{Bergner:2002we}.} In the thin-wall regime, this calculation can be performed analytically (see Ref.~\cite{Garbrecht:2015oea}). Specifically, we can treat the bubble wall as approximately planar, introducing a set of coordinates aligned parallel and perpendicular to the bubble wall (see Fig.~\ref{fig:planarwall}). In the limit $\mu R\gg1$, we can replace the discrete summation over angular momentum modes by an integral over a continuous momentum conjugate to the parallel coordinates $\mathbf{z}_{\parallel}$:
\begin{equation}
\frac{j(j+1)\hbar}{\mu^2R^2}\ \longrightarrow\ \frac{k^2}{\mu^2}\;.
\end{equation}
Making the change of variables $u^{(\prime)}\equiv\varphi(r^{(\prime)})/v$, the Green's function has the form
\begin{align}
G(u,u',m)\ &=\ \frac{1}{2\gamma m}\,\bigg[\vartheta(u-u')\bigg(\frac{1-u}{1+u}\bigg)^{m/2}\bigg(\frac{1+u'}{1-u'}\bigg)^{m/2}\nonumber\\&\times\bigg(1-3\,\frac{(1-u)(1+m+u)}{(1+m)(2+m)}\bigg)\bigg(1-3\,\frac{(1-u')(1-m+u')}{(1-m)(2-m)}\bigg)\:+\:(u\longleftrightarrow u')\bigg]\;,
\end{align}
where we have defined
\begin{equation}
m \ \equiv\ 2\bigg(1+\frac{k^2}{4\gamma^2}\bigg)^{1/2}\;.
\end{equation}

\begin{figure}
\centering
\includegraphics[scale=0.33]{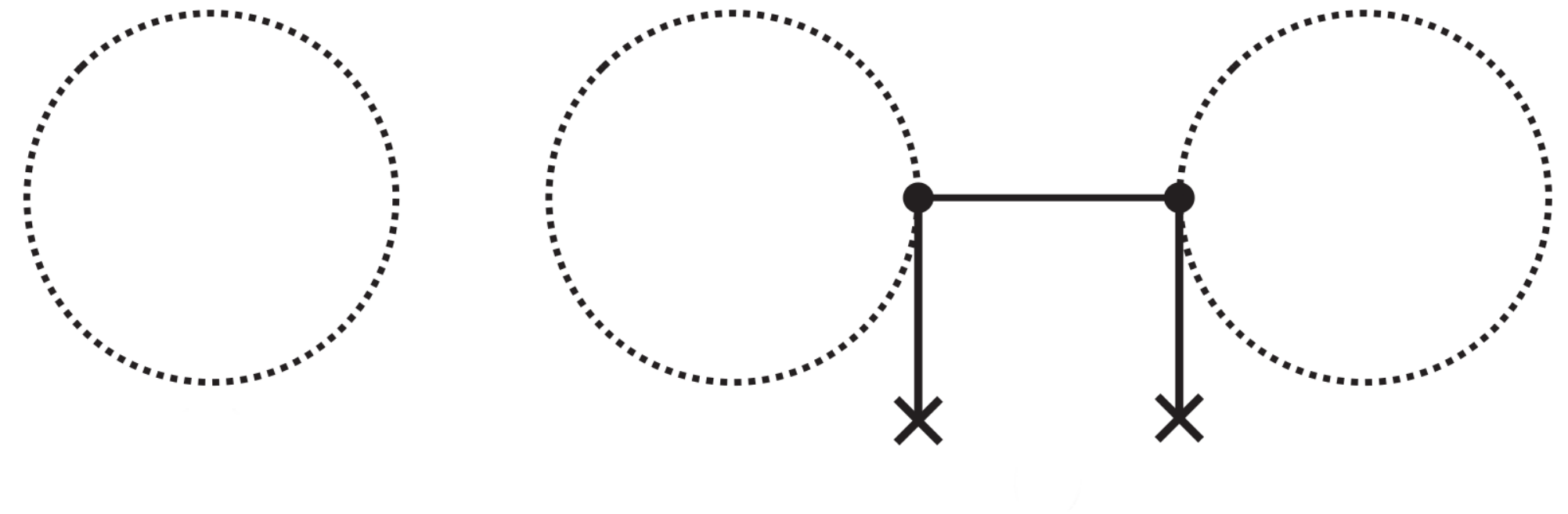}
\caption{\label{fig:Feyns}The dominant corrections to the tunneling rate at order $\hbar$ (left) and order $\hbar^2$ (right). Dotted lines are $X$ Green's functions, solid lines are $\Phi$ Green's functions and crosses correspond to insertions of the classical background field $\varphi^{(0)}$.}
\end{figure}

So as to make a meaningful perturbative truncation and to enhance the radiative effects, we supplement the model in Eq.~\eqref{eq:Lagrangian} with $N$ additional scalar fields:
\begin{equation}
\mathcal{L}\ \supset\ \sum_{i\,=\,1}^N\Bigg[\frac{1}{2}\big(\partial_{\alpha}X_i\big)^2\:+\:\frac{1}{2}\,m_{\chi}^2X_i^2\:+\:\frac{1}{4}\,\lambda\,\Phi^2\,X_i^2\Bigg]\;.
\end{equation} 
The dominant corrections in the $1/N$ expansion~\cite{tHooft:1973alw} of the tunneling rate are shown in Fig.~\ref{fig:Feyns}. In the limit $m_{\chi}^2\gg \gamma^2$, the renormalised $X$ tadpole is given by
\begin{equation}
\Sigma^R(u)\ =\ \frac{\lambda\gamma^2}{8\pi^2}\,\frac{\gamma^2}{m_{\chi}^2}\,\big[72+(1-u^2)(40-3u^2)\big]\;.
\end{equation}

\begin{figure}
\centering
\includegraphics[scale=0.45]{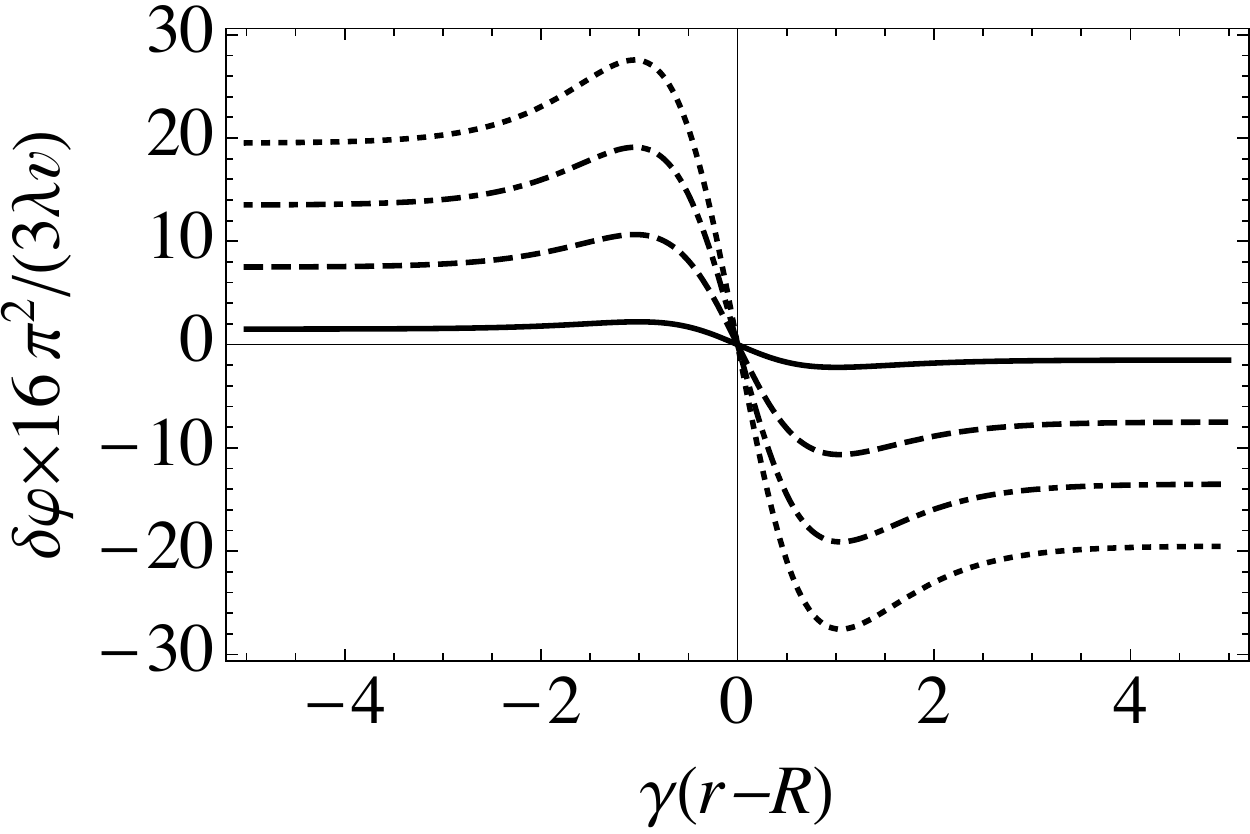}\hspace{2em}\includegraphics[scale=0.45]{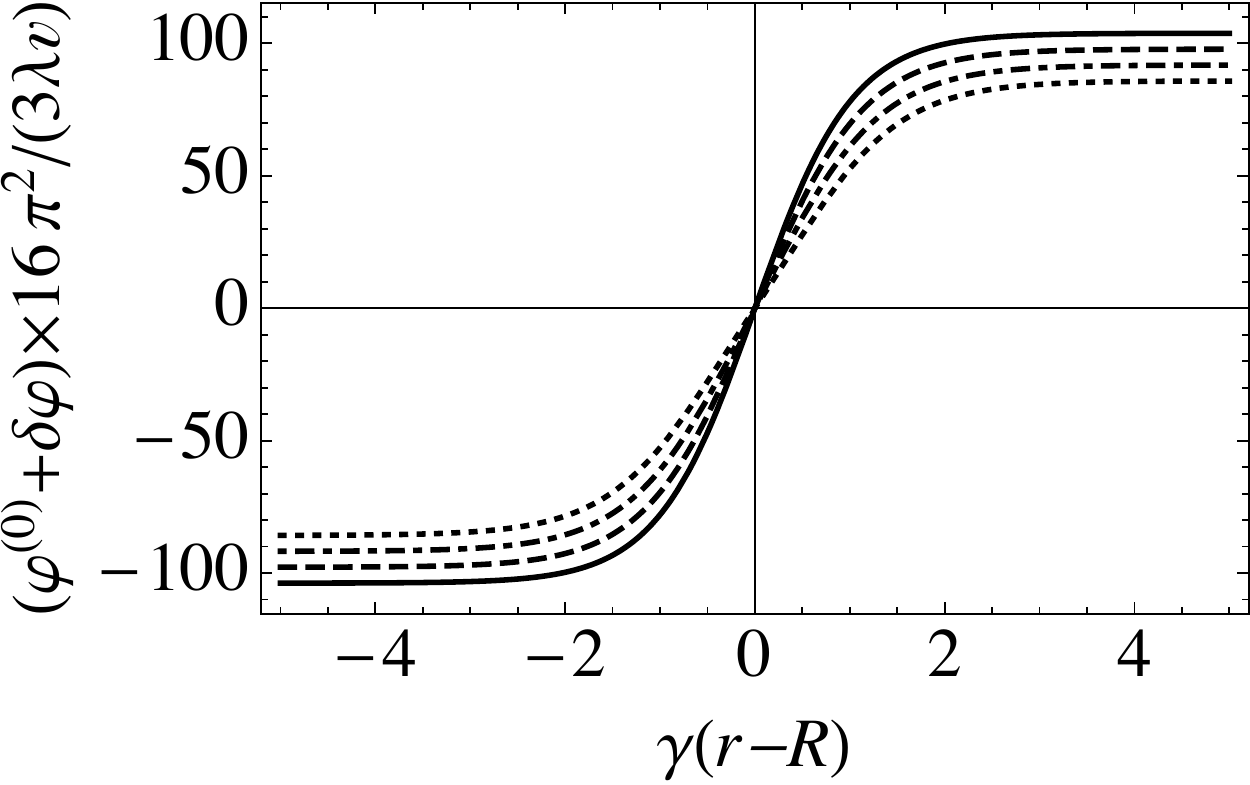}
\caption{\label{fig:deltaphi}The quantum correction $\delta\varphi$ (left) and quantum-corrected bounce $\varphi^{(1)}=\varphi^{(0)}+\delta\varphi$ (right) for $N\gamma^2/m_X^2$ equal to $0$ (solid), $0.5$ (dashed), $1$ (dash-dotted) and $1.5$ (dotted).}
\end{figure}

We can also calculate the quantum corrections to the bounce, which are given explicitly by
\begin{equation}
\delta\varphi(u)\ =\ -\:\frac{v}{\gamma}\int_{-1}^1{\rm d}u'\,\frac{u'G(u,u',m)|_{k\,=\,0}}{1-u^{\prime2}}\,\Big(\Pi^R(u')+N\,\Sigma^R(u')\Big)\;,
\end{equation}
where we have included the (subdominant) renormalised $\Phi$ tadpole ($\Pi^R$) for completeness:
\begin{equation}
\Pi^R(u)\ =\ \frac{3\lambda \gamma^2}{16\pi^2}\bigg[6+(1-u^2)\bigg(5-\frac{\pi}{\sqrt{3}}\,u^2\bigg)\bigg]\;.
\end{equation}
Figure \ref{fig:deltaphi} shows the quantum correction and quantum-corrected bounce for a range of values of $N\gamma^2/m_X^2$. The broadening of the bubble wall reduces the bounce action, leading to an increase in the tunneling rate. This is in qualitative agreement with the analysis of Ref.~\cite{Bergner:2003au}.

\section{Radiatively-generated instability}

\begin{figure}
\centering
\includegraphics[scale=0.45]{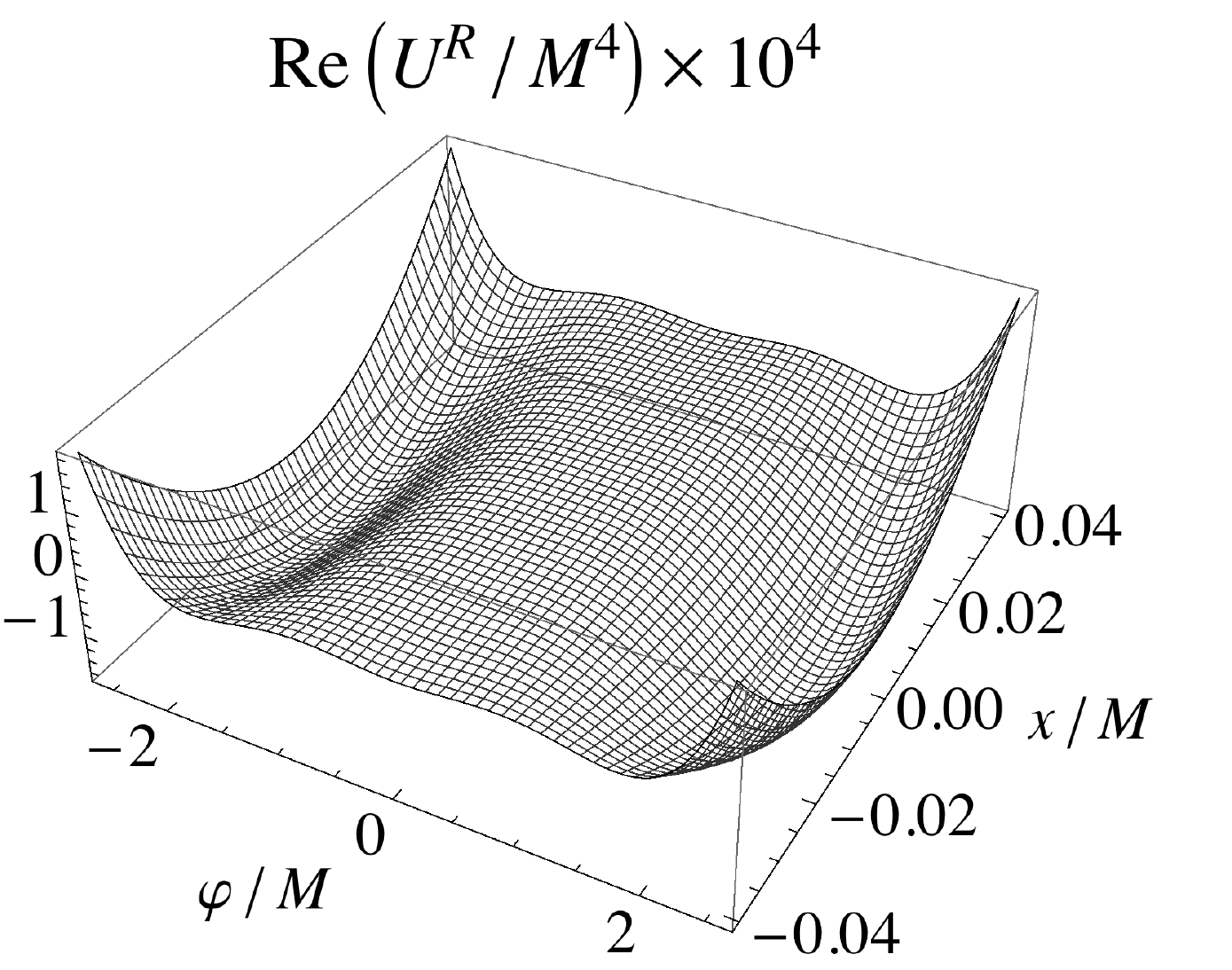}\hspace{2em}\includegraphics[scale=0.45]{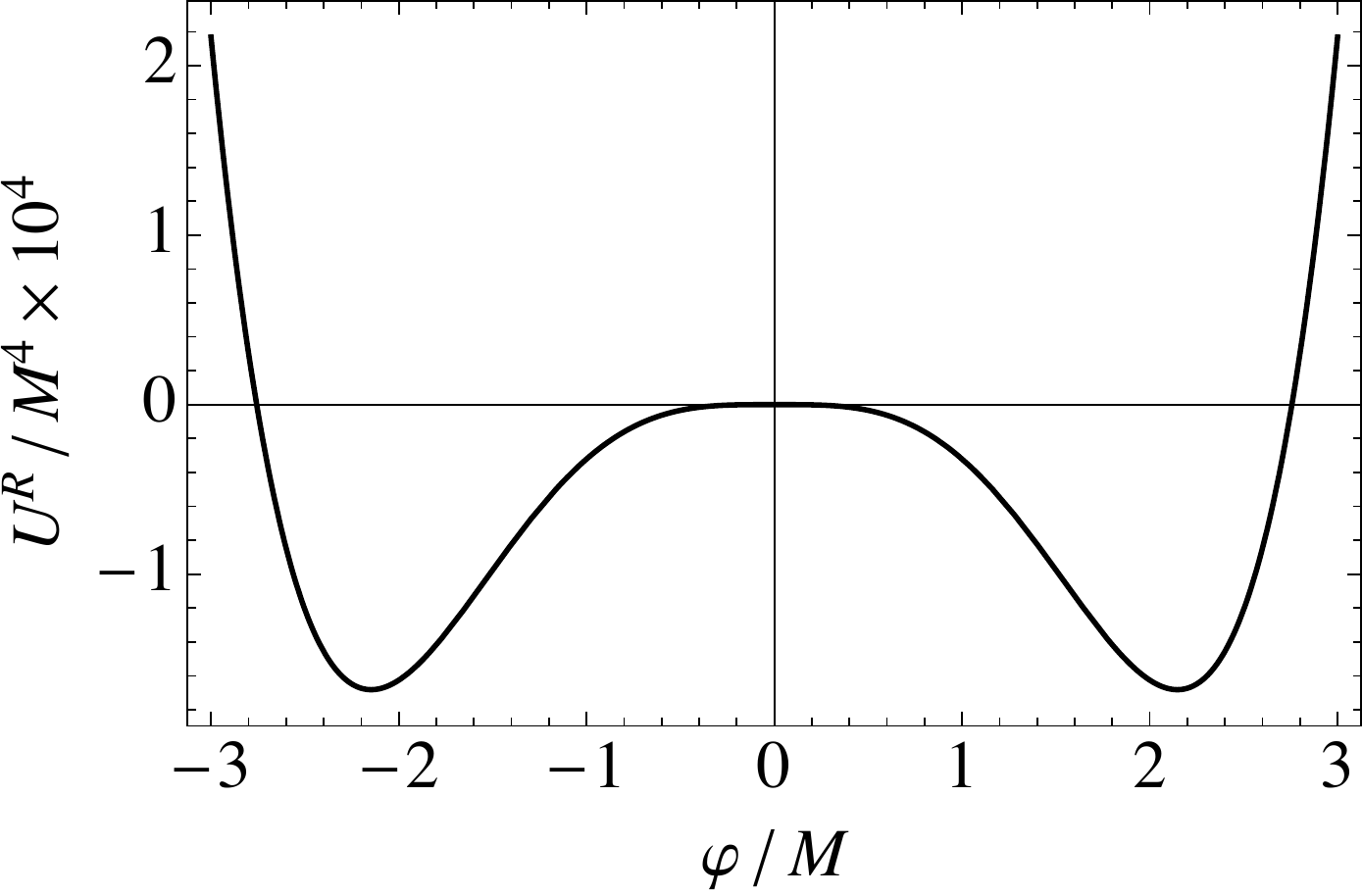}
\caption{\label{fig:CWplots}Three-dimensional plot of the renormalised effective potential in the unitary gauge for the model in Eq.~\eqref{eq:CWmodel} (left), as well as the cross-section in the plane $\chi=0$ (right).}
\end{figure}

We now turn our attention to a model where the instability instead arises at the quantum level. We start with a classically scale-invariant model (for $g=0$) with Lagrangian
\begin{equation}
\label{eq:CWmodel}
\mathcal{L}\ =\ \frac{1}{2!}\,\big(\partial_{\alpha}\Phi\big)^2\:+\:\frac{1}{2!}\sum_{i\,=\,1}^N\big(\partial_{\alpha}X_i\big)^2\:+\:\frac{1}{4}\,\lambda\,\Phi^2\sum_{i\,=\,1}^NX_i^2\:+\:\frac{1}{4}\,\kappa\,\sum_{i,\,j\,=\,1}^NX_i^2\,X_j^2\:+\:\frac{1}{3!}\,g\,\Phi^3\:+\:U_0\;.
\end{equation}
This model suffers logarithmic infra-red divergences at the one-loop level, which are regularised by introducing a scale $m$. This regulator mass is translated into a symmetry breaking scale $v$ via dimensional transmutation through the well-known Coleman-Weinberg mechanism~\cite{Coleman:1973jx}. Thus, at the one-loop level, the potential contains two minima, the degeneracy of which is broken by hand through the small cubic interaction. Choosing renormalisation conditions as in Ref.~\cite{Garbrecht:2015yza}, the renormalised Coleman-Weinberg (CW) one-loop effective potential is
\begin{equation}
U^R_{\mathrm{eff}}(\varphi)\ = \ \bigg(\frac{\lambda}{16\pi}\bigg)^2\,\varphi^4\Bigg[N\Bigg(\ln\frac{3\varphi^2}{\rho M^2}\:-\:\frac{3}{2}\Bigg)\:+\:F(\rho)\:+\:\frac{1}{6}\,g\,\varphi^3\:+\:U_0\Bigg]\:+\:\mathcal{O}(g^2)\;,
\end{equation}
where $\rho\equiv 6\kappa/\lambda$ and
\begin{equation}
F(\rho)\ \equiv\ \ln\,3\:+\:\frac{8}{(1-\rho^2)}\,\bigg(3\:+\:\rho\:+\:\frac{1+3\rho}{1-\rho}\,\ln\,\rho\bigg)\;.
\end{equation}
For $g\ll 32\pi^2v/(\lambda^2N)$, the global minima of the potential lie at
\begin{equation}
\varphi\ \approx\ \pm\,v\ =\ \pm\,\sqrt{\frac{\rho M^2}{3}}\,\exp\bigg[\frac{1}{2}+\frac{F}{2N}\bigg]\;.
\end{equation}
By using auxiliary fields to induce the symmetry breaking, we remain in the regime of validity of the one-loop approximation. The effective potential is plotted in Fig.~\ref{fig:CWplots}.

\begin{figure}
\centering
\includegraphics[scale=0.45]{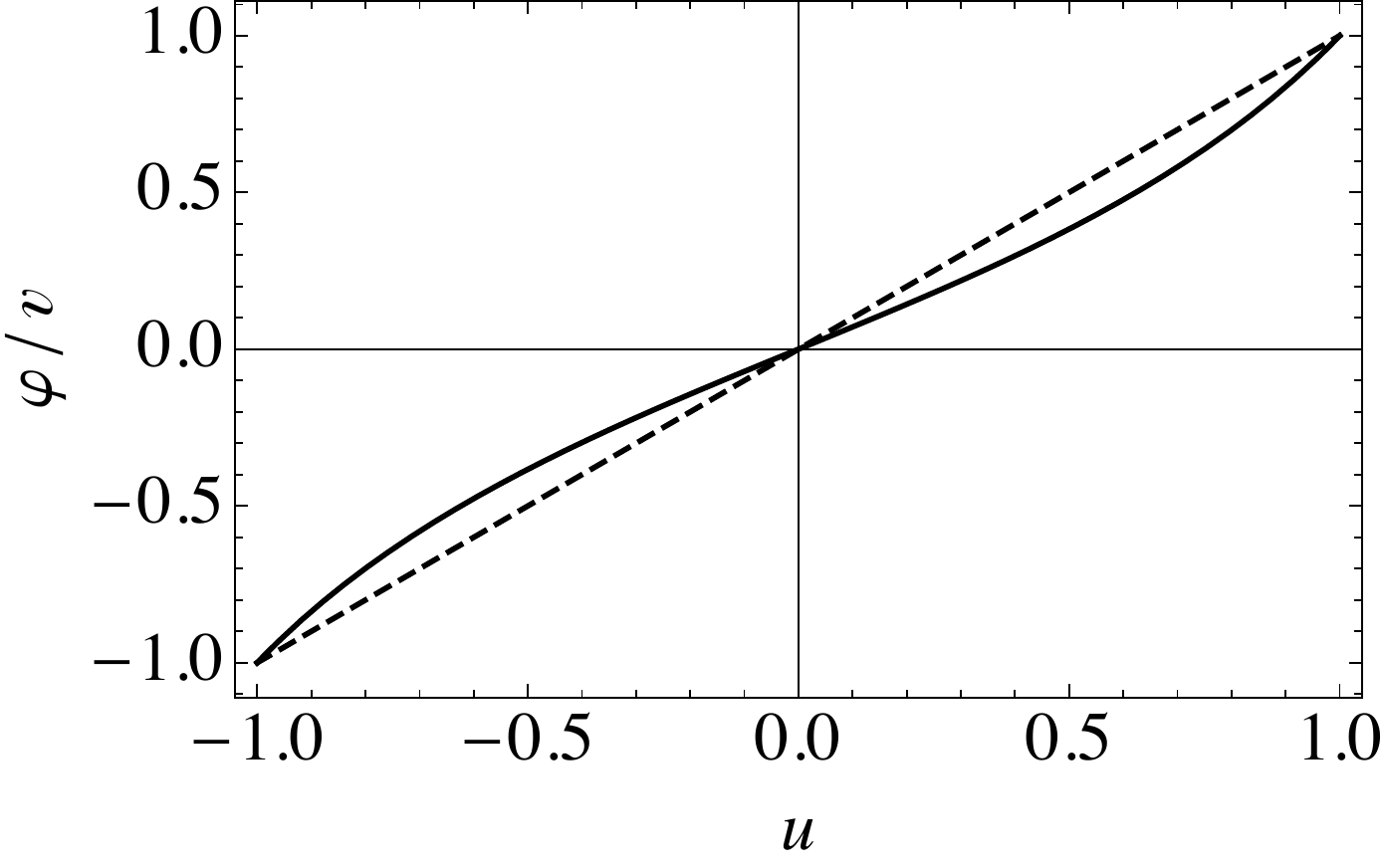}\hspace{2em}\includegraphics[scale=0.45]{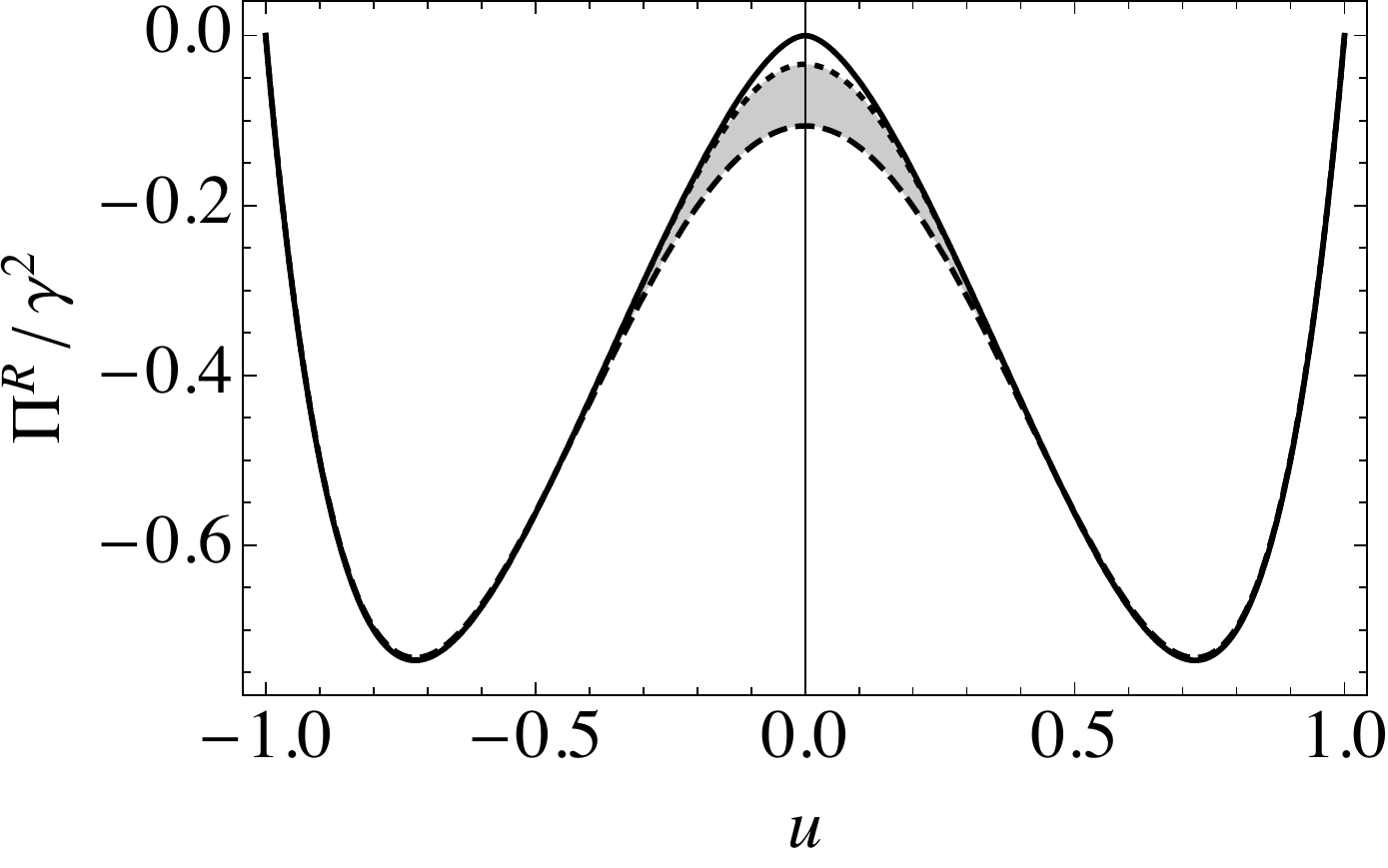}
\caption{\label{fig:selfcons}Left: The self-consistent bounce compared with the kink in Eq.~\eqref{eq:kink}. Right: The renormalised tadpole correction. The solid line is the CW (homogeneous) result, and the dotted and dashed lines are the self-consistent results for the first and last parameter points in the sampling range: $M=1$ with $0.04\leq \lambda^2 N\leq 0.4$.}
\end{figure}

For large $N$, the $1/N$ power counting is such that (i) we must treat the equation of motion for $\varphi$ at the one-loop level:
\begin{equation}
-\:\Delta^{(4)}\varphi(x)\:+\:\Sigma^R(\varphi;x)\varphi(x)\ =\ 0\;,\qquad \Sigma^R(\varphi;x)\ =\ \frac{\lambda N}{2}\,S(\varphi;x,x)\:+\:\delta\Sigma\;,
\end{equation}
where $S(\varphi;x,y)$ is the $X$ Green's function and $\delta\Sigma$ contains the relevant counterterms; (ii) we can treat the equation of motion for the $X$ field Green's function at tree-level:
\begin{equation}
\bigg(-\:\Delta^{(4)}\:+\:\frac{\lambda}{2}\,\varphi^2\bigg)S(\varphi;x,y)\ = \ \delta^{4}(x-y)\;;
\end{equation}
and (iii) the $\Phi$ Green's function decouples entirely. Working again in the thin- and planar-wall limits, we employ an iterative numerical procedure to determine the self-consistent bounce:
\begin{enumerate}
\item [1.] Calculate a first approximation to the bounce by promoting the homogeneous field configuration in the CW effective potential to a space-time dependent one:
\begin{equation}
-\Delta^{(4)}\varphi\:+\:U^{R\prime}_{\mathrm{eff}}(\varphi)\ =\ 0\;.
\end{equation}
\item [2.] Solve for the $X$ field Green's function.

\item [3.] Calculate the tadpole correction, renormalising in the homogeneous false vacuum.

\item [4.] Insert the tadpole into the quantum equation of motion and solve for the bounce.

\item [5.] Iterate over steps 2 to 5 until the solution has converged sufficiently.
\end{enumerate}
This procedure essentially undoes a gradient expansion, putting back the full dependence on the inhomogeneity of the solitonic configuration. The resulting self-consistent bounce and one-loop tadpole, as obtained in Ref.~\cite{Garbrecht:2015yza}, are shown in Fig.~\ref{fig:selfcons}.

\begin{figure}
\centering
\includegraphics[scale=0.45]{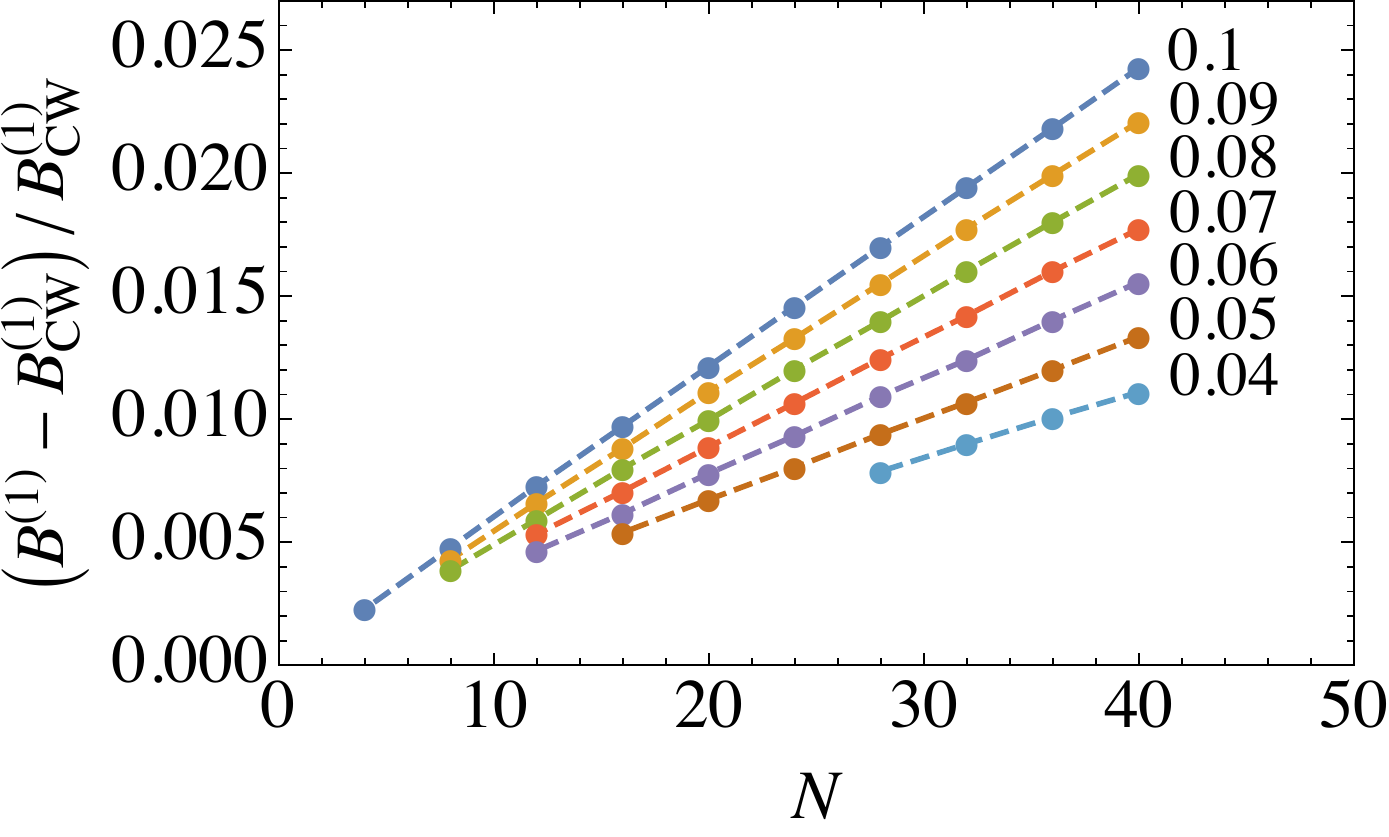}\hspace{2em}\includegraphics[scale=0.45]{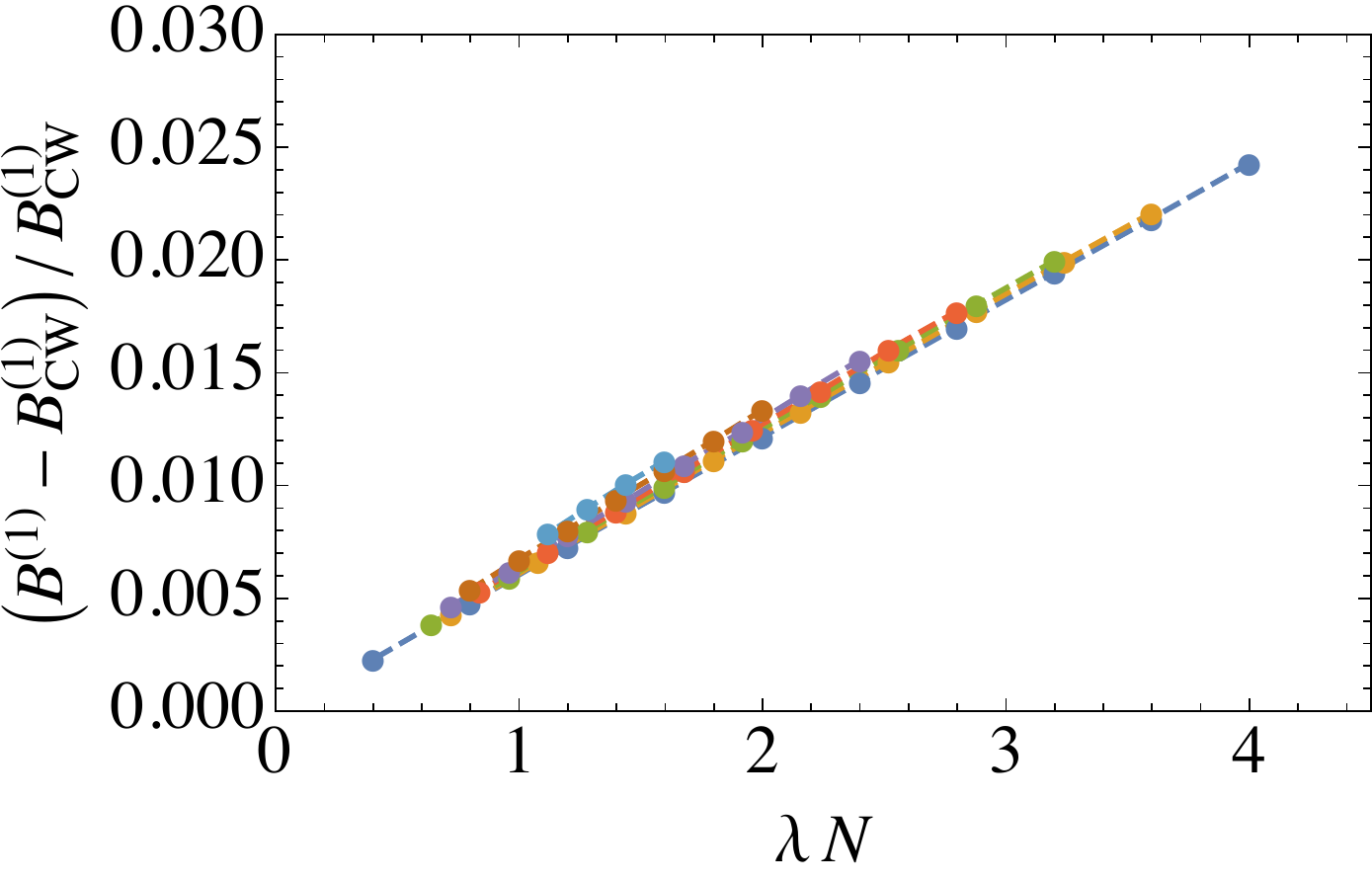}
\caption{\label{fig:B1}The fractional difference in the self-consistent fluctuation determinant $B^{(1)}$ relative to the Coleman-Weinberg result $B^{(1)}_{\rm CW}$ as a function of $N$ (left) and $\lambda N$ (right). The labels in the left-hand plot indicate values of $\lambda$.}
\end{figure}

The dominant dependence on the gradients occurs in the one-loop fluctuation determinant $B^{(1)}$. The results are shown in Fig.~\ref{fig:B1}. We see from the right-hand plot that the deviation between the self-consistent and (homogeneous) CW results scales like $\lambda N$ relative to the CW result. Scaling as $\lambda N$ times the homogeneous one-loop terms, the gradient effects can therefore compete with or even be more important than the two-loop effects, which scale as $\lambda\kappa N^2$ or $\lambda^2N$. This numerically confirms arguments presented in Ref.~\cite{Weinberg:1992ds}.

\section{Conclusions and future directions}

We have described a Green's function method for calculating the quantum corrections to the decay rate of false vacua that fully accounts for the background inhomogeneity of the solitonic, tunneling configuation. We have applied this technique to calculate the quantum-corrected decay rate analytically in the archetypal $\Phi^4$ theory and outlined a numerical procedure for the self-consistent calculation of the tunneling rate when the instability arises only at the quantum level. There, we have shown that the modification of the one-loop terms due to accounting fully for gradient effects can compete with the (homogeneous) two-loop terms. In the thin-wall regime, the gradient effects are suppressed due to the symmetry of the bounce solution about the mid-point of the bubble wall, and such a suppression is not anticipated in the thick-wall regime. A study of this regime is of particular relevance to the instability of the SM electroweak vacuum and will be presented elsewhere.

\ack

The work of PM is supported by STFC grant ST/L000393/1. BG acknowledges the support of the Deutsche Forschungsgemeinschaft (DFG) through the cluster of excellence `Origin and Structure of the Universe'.  PM would like to thank Wen-Yuan Ai for on-going collaboration, the organisers of DISCRETE2016 for their hospitality and the opportunity to present this work, and the participants for their questions, comments and suggestions.

\end{document}